# Top and side gated epitaxial graphene field effect transistors

Xuebin Li[1], Xiaosong Wu[1], Mike Sprinkle[1], Fan Ming[1], Ming Ruan[1], Yike Hu[1], Claire Berger[*, 1, 2] and Walt A. de Heer[1]

1  Georgia Institute of Technology - School of Physics, Atlanta, GA-30332, USA
2  CNRS- Institut Néel, BP 166, 38042 Grenoble Cedex 9, France
*  Corresponding author

Three types of first generation epitaxial graphene field effect transistors (FET) are presented and their relative merits are discussed. Graphene is epitaxially grown on both the carbon and silicon faces of hexagonal silicon carbide and patterned with electron beam lithography. The channels have a Hall bar geometry to facilitate magnetoresistance measurements. FETs patterned on the Si-face exhibit off-to-on channel resistance ratios that exceed 30. C-face FETs have lower off-to-on resistance ratios, but their mobilities (up to 5000 cm$^2$/Vs) are much larger than that for Si-face transistors. Initial investigations into all-graphene side gate FET structures are promising.

**1 Introduction** Epitaxial graphene on silicon carbide (EG) is has demonstrated its importance for fundamental graphene research and its applications potential [1, 2]. Continuous layers can be grown over SiC wafer-size surfaces, thereby allowing large scale patterning by conventional lithographic methods as has recently been demonstrated[3]. Recently discovered multilayer epitaxial graphene (MEG) shows large electronic mobilities [2, 4], with values exceeding 250,000 cm$^2$/Vs at room temperature. Due to the rotational stacking, the layers are effectively decoupled and their properties are similar to a single graphene sheet [5, 6]. A major challenge is to develop electrostatic gating schemes to control the charge density of the material without significantly affecting the carrier mobilities of the pristine material [3, 7, 8]. We show that epitaxial graphene grown on both Si-terminated and C-terminated faces can be gated by electrostatic gates patterned on top of the graphene. We further show that significant gating can be achieved in an all graphene, side-gated structure. This is a particularly promising scheme that potentially overcomes the detrimental effect of gate dielectrics on the graphene layer.

**2 Graphene FET preparation and structure** Epitaxial graphene on hexagonal SiC (4H or 6H) is produced by thermal desorption of silicon from a hexagonal silicon carbide [2]. EG produced in ultrahigh vacuum results in low quality ultrathin multilayer films or single graphene layers. The quality of the graphene films on both the silicon and the carbon terminated faces is dramatically improved [9]. using an inductively heated vacuum furnace at temperatures ranging from 1400C-1600C. Growth and structural details have been discussed elsewhere [2, 9]. It is important to note that AFM, STM and X-ray confirm that for both the C- and the Si-face, the graphene films produced at high temperature in the induction furnace have a far lower defect density than films produced in an UHV environment.

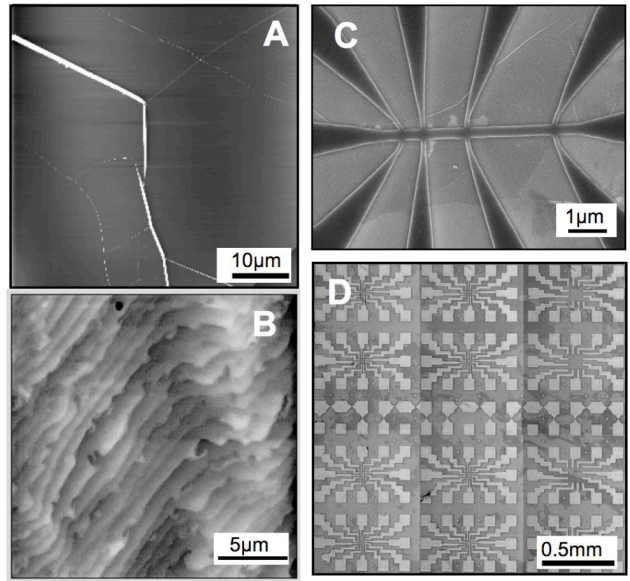

**Figure 1** Typical AFM images of furnace grown graphene layers on a C-face (A) and Si-face (B). (A) : the C-face multilayered epitaxial graphene is flat over several tens of micron large area. The layers are continuous, draping over the entire surface of the sample. The white transversal lines are graphene puckers showing the continuity of the graphene films. (B) few graphene layers drape over the SiC step structure (C): SEM picture of patterned Hall bar structure. The ribbon is patterned on a single terrace, with graphene pads extending out towards the Pd/Au contacts. (D) Example of integrated structures on a SiC chip, featuring a pattern of a hundred ribbons. The background contrast is an artifact from the tape on the back of the transparent SiC chip.

Silicon face films have from 1 to about 5 graphene layers. The graphene layer at the SiC/graphene interface acquires a negative charge density $n_0 \sim 5 \times 10^{12}$ electrons/cm [1,10-12]. The first graphene layer on the carbon face is also charged [2, 13] : $n_0 \sim 5 \ 10^{12}$ electrons/cm$^2$. In contrast to the silicon face however, the stacking of the subsequent layers (1- 100 layers) is a non-graphitic commensurate rotated phase [5], which preserves the integrity of the massless Dirac particles, as experimentally demonstrated in Landau level spectroscopy [4], in magneto scanning tunneling spectroscopy [14], Raman spectroscopy [15], and in ARPES experiments. The fields at the interface are screened after about two layers [16], and subsequent layers are essentially uncharged [4], ($n<5 \times 10^9/cm^2$), explaining why transport in (un-gated) MEG samples is primarily through interface graphene layer [2].

Fig. 1 A and B shows typical AFM images of graphene on the C- and Si-faces, respectively. The parallel steps on the Si face (Fig. 1B) correspond to approximately 1nm steps on underlying SiC substrate over which the continuous graphene layers are draped. On the C-face (Fig 1A) very large domains are observed of several tens of microns. These domains are atomically flat. The rms roughness measured[9] with X-ray diffraction is instrument- limited, i.e. less than 0.05nm The structural coherence length of graphene on the C-face as determined by x-ray diffraction is essentially limited by the terrace width and not by the graphene grain size.

Thermal contraction of SiC and graphene during cooling produces puckers on the graphene film causing the white lines observed in Fig 1A. The puckers traverse the SiC steps indicating that the graphene layers are continuous over the SiC steps and extend over very large areas in agreement with STM, STS, X-ray diffraction and LEEM measurements. In fact no break in the graphene top layer on the C-face has ever been observed by STM which suggests that at least this layer and possibly others as well, span the entire 3.5 X 4.5 mm size chip. From ellipsometry (and confirmed by LEEM) the film thickness is found to vary within ± 2 graphene layers on the carbon face (with typically 10 layers) and ± 1 layer on the Si-face (with typically 3 layers) over the entire chip surface. Reported contrast variations observed in optical microscopy of MEG samples[3] actually result from interference effects at the interface and are not due to thickness variations. MEG thickness variations measured by LEEM indeed do not correlate with the optical contrast.

The top-gated epitaxial graphene transistors consist of graphene ribbons in a Hall bar configuration supplied with a gate stack consisting of a spin-coated dielectric layer that is coated with aluminum. The FETs are produced as follows. Metal contacts (Pd/Au) are evaporated on a graphitized silicon carbide chip (3.5 mm X 4.5 mm) through a shadow mask. The sample is subsequently e-beam patterned (JEOL JBX- 9300FS, 100 keV, 2 nA) using hydrogen silsesquioxane (HSQ) or PMMA e-beam resist. The graphene layers are oxygen plasma etched to produce a graphene ribbon that defines the FET channel. Next, a Pd/Au layer is deposited on HSQ to produce the gate structure. Fig. 1C shows a SEM image of a completed structure. The pattern can be repeated for multiple structure integration as seen for instance in Fig.1D.

Transport measurements are performed in the four-probe configuration with a low frequency lock-in detection and dc gating. We present here results on patterned Hall bars on a Si-face film (S1: 3.5μmx12.5μm, N~3-4 layers), a C-face (C1: 3.5μm x 12μm with about 10 graphene layers), and a double side-gate C-face sample (0.1μm x 1.3μm).

### 3 Graphene FET on Si-face of SiC

**3.1 Si-face of SiC** Fig. 2B shows the square resistance $\rho_{xx}$ as a function of gate voltage $V_g$, at room temperature of a top-gated Si-face FET (sample S1). The resistance peaks at 36 kΩ/square for a gate voltage of $V_g max = -3.5V$, due to the charge on the graphene layer $n_0$. A resistance change of a factor of 30 is obtained for a gate voltage change of 6 V and represents the largest on-off current ratio measured for epitaxial graphene FETs (it is comparable to measurements on exfoliated graphene FETs [17, 18]). The conductivity σ plotted vs $V_g$ in Fig. 2A is symmetric around the minimum $G_{min}$ as previously observed for epitaxial graphene [3] and qualitatively similarly to exfoliated graphene FETs [17, 18]. However the minimum $G_{min} \sim 0.75\ G_0$ (where $G_0 = e^2/h$) is considerably smaller than previously observed and rather close to the theoretically predicted value [19] $G_{min} = 2/\pi\ G_0 = 0.63\ G_0$, as also found in optical absorption spectra of epitaxial graphene [20]. Note that exfoliated graphene FETs exhibit minimum conductivities [18] of about $4G_0$. For those FETs the large values have been attributed to large charge density variations ("electron-hole puddles") on or under the graphene layer [21].

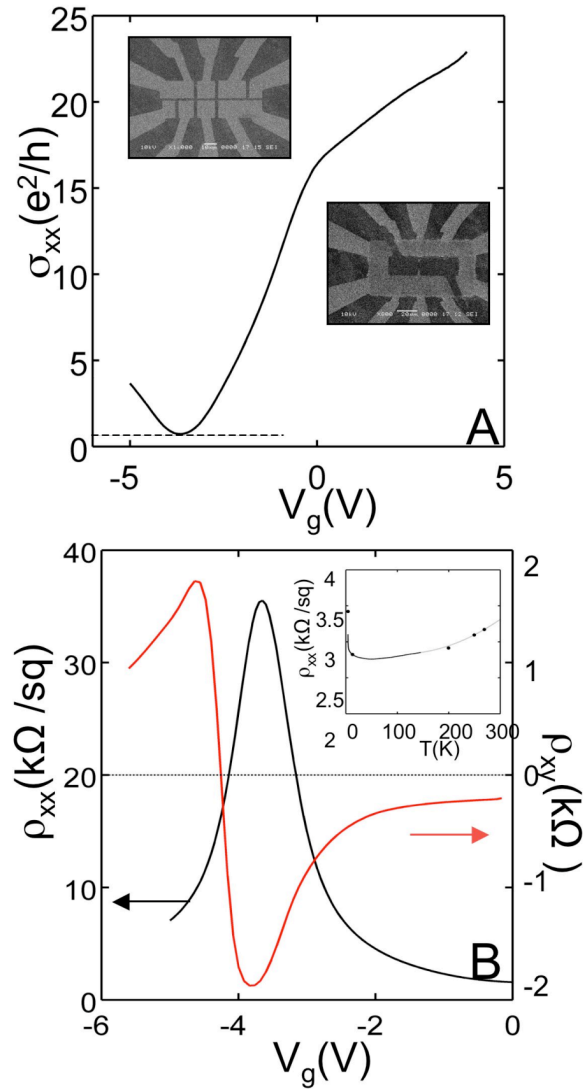

**Figure 2** (A) Conductivity $\sigma_{xx}$ as a function of gate voltage at 300K for graphene on Si-face SiC (Hall bar 3.5 μm x 12.5μm). The ratio of maximum to minimum resistance is $I_{on}/I_{offf}$=31. The minimum conductivity is close to the theoretical one $2e^2/\pi h$ (dotted line). Inset: the measured top gated ribbon, before and after gate deposition (spin-on HSQ resist and evaporated metal gates on top). (B) Resistivity $\rho_{xx}$ and Hall resistance $\rho_{xy}$ as a function of gate voltage at 5 Tesla and 300K The resistivity peaks when $\rho_{xy}$ changes sign. Inset: temperature dependence of $\rho_{xx}$ .(dotted line: interpolation between data point (dots)).is the figure caption.

Similar resistance behavior is observed in a magnetic field of 5T. As seen in Fig. 2B, the sign reversal of $\rho_{xy}$ at Vg~-3.5 V indicates at the crossover from electron to hole doping. From the Hall coefficient the charge density of sample S1 at $V_g$=0 is found to be $n_H$=8x10$^{12}$ cm$^{-2}$. This value is consistent with ARPES measurements of monolayer Si face graphene films [10, 11]. The room temperature Hall mobilies μ ~500 cm$^2$/Vs, which are in the range of mobility measured previously on top-gated Si-face graphene FETs [3]. Higher Hall mobilities (~1600 cm$^2$/Vs) are observed in pristine Si-face films [1].The mobilities are insensitive to temperature, implying that the mobility is dominated by impurity scattering caused by the dielectric on the graphene layer.

**3.2 C-face of SiC** We next turn to C-face, MEG transistors. Sample C1 is patterned on a C-face 10 layer MEG film. The FET consists of a Hall bar that is supplied with 3 gates along the 3.5μm x 12μm channel. From Hall measurements the charge density n=$10^{13}$ electrons/$cm^{-2}$ and the device Hall mobility μ ~1000 $cm^2$/Vs, and indicating a large reduction compared with the pristine material (μ >10,000 $cm^2$/Vs).

Like Si-face FETs, the C-face FETs exhibit a resistance minimum at the charge neutrality point of the graphene layer (or layers) that are affected by the gate. However, in contrast to Si face FETs for which the minimum conductivity always occurs for negative $V_g$, the minimum conductivity for C face FETs are observed for positive or negative $V_g$ as seen previously [3] and here in Figure 3 for gates positioned along the same ribbon. The ambiguous polarity of the minimum indicates that the HSQ gate dielectric introduces charges on top graphene layer, as also observed elsewhere [22, 23]. Furthermore, the graphene film is relatively thick. Consequently, because of screening, the gate potentials affect only the top few layers and not the (charged) interface layer. Therefore, the highly charged interface layer essentially serves as conductor in parallel with the transistor channel and provides a constant, device specific, offset in the conductivity. This offset varies from one device as has been observed before [3]. This effect is currently a serious limitation for C-face FETs. Note however that room temperature mobilities larger than 250,000 $cm^2$/Vs have been measured on C-face graphene layers, and FET from C-face can have FET mobilities μ>5000 $cm^2$/Vs, as measured previously [3]. Whereas the mobilities for C-face graphene is larger than for Si-face graphene, unfortunately the $I_{on}/I_{off}$ ratios are significantly lower due to the shorting effect of the interface layer.

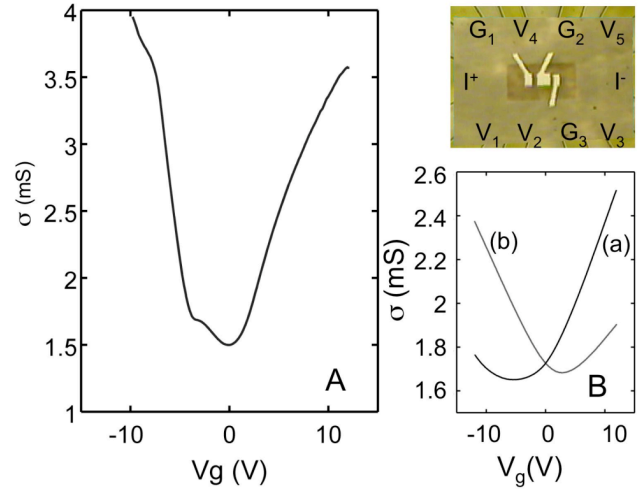

**Figure 3** Conductivity as a function of gate voltage for a C-face multilayer epitaxial graphene Hall bar, and optical composite image of the gated structure. Three gates (G1, G2 and G3, light color) evaporated on top of the dielectric (light brown rectangle) cover partially the ribbon laying between the current leads I. (A): conductivity for a portion of the ribbon entirely covered by the gate (gate G1, voltage probe V1 and V2). (B) ribbon partially gated (voltage probes V1 and V3, (a) gate G1, (b) gate G2). As expected a larger modulation ($I_{on}/I_{off}$ ~3) is observed in (A). Depending on the conditions, the gated portion of the ribbon can be p (a) or n (b) doped.

The low intrinsic on-off ratios in graphene FETs are a direct consequence of the semi-metallic nature of graphene, whose density of states increases linearly with energy away from the charge-neutrality point (the "Dirac-point"). Moreover, the conductivity at the Dirac point does not vanish but saturates at its theoretical value[19] of $2/\pi \, e^2/h$ ~ 24 μS, causing a significant off current. Much greater on-off ratios can be obtained in narrow ribbons due to the bandgap that opens. The bandgap is predicted and has been observed to be inversely proportional to the ribbon width. This effect can be exploited in device structures, even though its full potential can only be realized in sub-10 nm ribbons, for which the band gap >0.1 eV. Sub-10 nm graphene ribbons still pose a formidable technological challenge and is difficult to achieve even with state of the art e-beam lithographers.

**3.3 Side gate** Fig. 4 shows a specific FET configuration that is particularly promising. It is an all-graphene transistor that is side-gated. The advantages are clear: the FET is produced in a single etching step that at the same time defines the channel and the side gates. Moreover, it does not require a dielectric on top of the graphene thereby eliminating some of the problems. A resistance modulation of 22% is seen for a 1 V change in Vg. The properties of this FET are encouraging however not yet compelling. Only a relatively small response was achieved and there was significant gate-to-channel leakage. Improved designs are currently investigated.

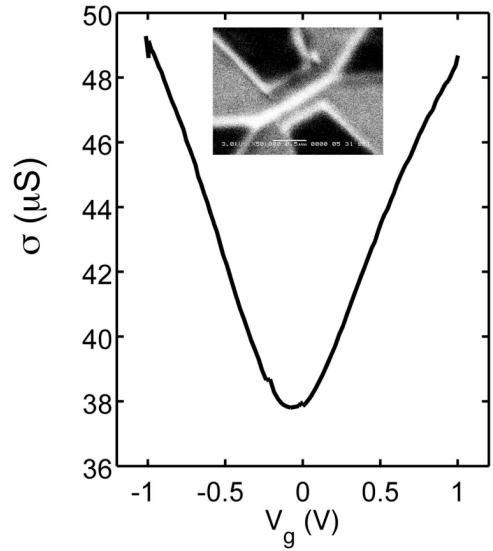

**Figure 4** Conductivity for a C-face multilayer epitaxial graphene nano-ribbon (width 50 nm) with split gates, and SEM image of the gated structure.

**4 Summary** In summary, first generation field effect transistors patterned on epitaxial graphene on silicon carbide are promising. Continuous epitaxial graphene layers can be routinely grown over large areas. Patterning is performed using standard lithographic methods. While FETs patterned on the Si-face typically exhibit larger on-off ratios than those patterned on the C-face, the mobilities of the latter (l~ 5000 $cm^2$/Vs) are about an order of magnitude larger than the former. This is a direct consequence of the relatively thicker layer on the C-face (5-10 sheet) and the relatively large charge density on the interface layer compared to the thin (few sheets) Si face material. Initial investigations of narrow ribbons are encouraging and C-face channels as narrow as 15 nm have been produced with Ion/Ioff ratios of about 10. Prototype all-graphene side-gate structures are also promising. It should be mentioned that despite the poor characteristics, epitaxial graphene transistors have already been shown to operate in the GHz frequency range[24].

Future investigations center around producing better gate dielectrics, establishing better lithography methods and producing thinner C-face material and passivation of the interface. Despite these formidable challenges, epitaxial graphene is currently the only graphene material with demonstrated large-scale integration potential.

**Acknowledgements** We acknowledge financial support from the W. M. Keck Foundation, NSF (NIRT grant #4106A68), and a travel grant from the French Partner University Fund. This work is part of the NSF-MRSEC at Georgia Tech (grant # 0820382).